\documentclass{icrc2009}

\usepackage{graphicx}   
\usepackage[caption=false]{caption}    
\usepackage[font=footnotesize]{subfig} 
\usepackage{fixltx2e}
\usepackage{url}

\newcommand{\shorttitle}[1]%
{\markboth{Proceedings of the 31\MakeLowercase{$^{st}$} ICRC, {\L}\'{o}d\'{z} 2009}{#1} }
\newcommand{\etal}{\MakeLowercase{\textit{et al. }}} 

\begin{document}
\title{Simultaneous multi-frequency observations of PG\,1553+113}

\author{\IEEEauthorblockN{Nijil Mankuzhiyil\IEEEauthorrefmark{1},
			  Daniela Dorner\IEEEauthorrefmark{2},
                          Elisa Prandini\IEEEauthorrefmark{3},
                          Massimo Persic\IEEEauthorrefmark{4},
                          \\                          
                          \\
                          for the  MAGIC collaboration,
                          \\
                          \\
                          Elena Pian\IEEEauthorrefmark{5},
                          Filippo D'Ammando\IEEEauthorrefmark{6} and
                          Stefano Vercellone\IEEEauthorrefmark{7}
                          \\
                          \\
                          for the AGILE collaboration}

                           \\

\IEEEauthorblockA{\IEEEauthorrefmark{1}University of Udine and INFN Trieste, Italy}

\IEEEauthorblockA{\IEEEauthorrefmark{2}ETH Zurich, Switzerland}

\IEEEauthorblockA{\IEEEauthorrefmark{3}University of Padova and INFN Padova, Italy}

\IEEEauthorblockA{\IEEEauthorrefmark{4} INAF and INFN Trieste, Italy}

\IEEEauthorblockA{\IEEEauthorrefmark{5} INAF Osservatorio Astronomico Trieste, Italy}

\IEEEauthorblockA{\IEEEauthorrefmark{6}INAF-IASF Roma, Italy}

\IEEEauthorblockA{\IEEEauthorrefmark{7}IASF Milan, Italy}

}

\shorttitle{N. Mankuzhiyil \etal simultaneous mwl observation of PG 1553}
\maketitle

\begin{abstract}
We report simultaneous multi-frequency observations of the blazar PG\,1553+113, that were 
carried out in March-April 2008. Optical, X-ray, high-energy (HE: $\geq 100$\,MeV) $\gamma$-ray, and 
very-high-energy (VHE: $\geq 100$\,GeV) $\gamma$-ray data were obtained with the KVA, REM, {\it Rossi}XTE/ASM, 
AGILE and MAGIC telescopes. This is the first simultaneous broad-band (i.e., HE+VHE) 
$\gamma$-ray observation of a blazar.  The source spectral energy distribution derived 
combining these data shows the usual double-peak shape, and is interpreted in the framework 
of a synchrotron-self-Compton model.

  \end{abstract}

 
\section{Introduction}
The transformation  of gravitational energy from an accreation disk around a supermassive 
black hole into radiation is commonly believed to be the underlying cause of emission in Active 
Galactic Nuclei (AGNs). Furthermore, much of the emission is relativistically beamed perpendicular 
to the disk by a mechanism that, although not completely understood yet, is most likely related to 
the focusing properties of the rotating, fully-ionized accretion disk \cite {BlandfordZnajek1997}. 
It is believed that the viewing angle of the observer determines the observed phenomenology of 
AGNs \cite{UrryPadovani1995}. The AGNs whose relativistic plasma jets point towards the observer are 
called blazars. The blazar class includes Flat-Spectrum Radio Quasars (FSRQs) and BL\,Lac objects, 
the main difference between the two classes being their optical emission lines, which are strong 
and quasar-like for FSRQs and weak or absent in BL Lacs.

The overall (radio through $\gamma$-rays) spectral energy distribution (SED) of blazars usually 
shows two broad non-thermal continuum peaks. For high-energy-peaked BL\,Lac objects (HBLs), the 
first peak is in the UV/soft-X-ray bands [as opposed to IR/optical for low-energy-peaked BL\,Lac 
objects (LBLs)] whereas the second peak is in multi-GeV band (multi-MeV for LBLs). The low-energy 
peak is most commonly believed to be synchrotron radiation from a non-thermal population of electrons 
moving in a tangled magnetic field threading the emitting plasma region -- the latter is modeled 
as a sperical blob that moves relativistically along the jet. The second peak forms by Compton 
upscattering of lower-energy photons by the same population of electrons that is responsible for 
the synchrotron emission. The lower-energy photons can originate either from the synchrotron radiation 
constituting the low-energy bump (Synchrotron Self Compton: SSC), or from outside the relativistic 
plasma blob (External Compton: an external source of 'seed' photons could be the accretion disk 
\cite{Dermer1993} and/or broadline region \cite{Sikora1994}). 

Blazars often show violent flux variability, that may or may not appear correlated in the various 
energy bands. Simultaneous observations are then crucial to understand any patterns of temporal 
variability as well as the emission processes. 

The high-energy-peaked BL Lac source PG\,1553+113 was firmly detected at $E \geq 200$\,GeV 
by MAGIC at an 8.8$\sigma$ significance level, based on the data from 2005 and 2006 \cite{Albert2007}. 
The source had also been tentatively detected at VHEs by H.E.S.S., 
at a 4$\sigma$ significance level (5.3$\sigma$ using low-energy threshold analysis: 
\cite {Aharonian2006}), which was confirmed later using 2005 and 2006 data \cite{Aharonian2009}.
The lack of detection of spectral lines (neither in emission nor in absorption) in the optical 
spectrum of PG\,1553+113 makes it impossible to measure its redshift directly \cite{FalomoTreves1990}. 
However, an ESO-VLT spectroscopic survey of unknown-redshift BL\,Lac objects suggests $z>0.09$ 
\cite{Sbarufatti2006}. On the other hand, the absence of a break in the intrinsic VHE $\gamma$-ray 
spectrum may sugges $z<0.42$ \cite{MazinGoebel2007}. \\

\section{VHE $\gamma$-rays : MAGIC observations}
 
The MAGIC Telescope \cite{Baixeras2004}, \cite{Cortina2005} is a latest 
generation Imaging Atmospheric Cherenkov Telescope (IACT) at La Palma, Canary 
Islands, Spain (28.3$^{\circ}$N, 17.8$^{\circ}$W, 2240~m a.s.l.). Thanks to its
low threshold of 50\,GeV \cite{Albert2008a}, MAGIC is well suited for multi-frequency 
observations, together with the instruments operating in the GeV range. The 
parabolic shaped reflector, resulting in a total mirror area of 236\,m$^2$, allows 
MAGIC to collect Cherenkov light from particle showers initiated by gammas or other 
particles in the atmosphere and focus it onto a multi-pixel camera, composed of 577  
photomultipliers. The total field of view of the camera is 
3.5$^\circ$. The incident light pulses are converted into optical signals and 
transmitted via optical fiber to a 3-level trigger system.  The selected 
events are digitized by 2~Gsamples/second Flash ADCs \cite{Goebel2007}. With a 
statistical analysis of the recorded light distribution and the
orientation of the shower image in the camera, the energy of the primary
particle and its incoming direction are reconstructed. 

The MAGIC observations used in this paper were performed on 16-18 March and 13, 
28-30 April 2008. The zenith angle of the data set ranges from 18 to 36 degrees. 
Observations were performed in wobble mode \cite{Fomin1997}, where the object 
was observed at 0.4 degree offset from the camera center in opposite directions 
every 20 minutes. After the quality selection of the data the total effective 
observation time is 7.18 hours. 

An automatic analysis pipeline (see \cite{Dorner2005}) was used to process the data. 
After an absolute calibration with muons \cite{Goebel2005}, and an absolute mispointing correction, noise
subtraction and background reduction is achieved using the charge distribution and the arrival
time information of the pulses of the pixels \cite{Aliu2009}. Three OFF regions, defined in the 
same data set symmetrically to the ON region with respect to the camera center, were used to 
determine the remaining background. To select the gamma-like events, a dynamical cut in Area
(Area=$\pi\cdot$WIDTH$\cdot$LENGTH) versus SIZE and a cut in
$\vartheta$ were applied \footnotemark. More details of the cuts can be found in \cite{RiegelBretz2005}. 
\footnotetext{Width, Length and Size are the image parameters in \cite{Hillas1985}.}  

For the calculation of the differential energy spectrum, a loose cut in Area was selected to 
ensure that more than 90\% of the MC gammas after the image cleaning survive the cut. To study 
the dependence of the spectral shape on the cut efficiency, a different cut in Area with cut 
efficiencies of between 50\% and 95\% for the entire energy range was applied. The gray area 
in Fig.1 is the result of this study. For data affected by calima (i.e.\ windblown sand dust 
from Sahara in an air layer between 1.5\,km and 5.5\,km~a.s.l. causing absorption of the Cherenkov 
light), a correction has been applied. The method is described in detail in \cite{Dorner2009}.

\section{HE $\gamma$-rays: AGILE observations}

The Gamma-Ray Imaging Detector (GRID: 30\,MeV--30\,GeV) on board the spaceborne HE-$\gamma$-ray telescope 
AGILE
\footnote{  Astro-rivelatore Gamma a Immagini LEggero; 
		\cite{Tavani2008a}, \cite{Tavani2008b}.} 
observed PG\,1553+113 in three different time periods: 16-21 March, 25-30 March and 10-30 April 2008. 
The GRID data were analyzed using the AGILE standard pipeline (see \cite{Vercellone2008} for a detailed 
description of the AGILE data reduction), with a bin size of $0.25^{\circ} \times 0.25^{\circ}$ 
for E $>$ 100 MeV. Only events flagged as confirmed $\gamma$-rays and not recorded while the 
satellite crossed the South Atlantic Anomaly were accepted. We also rejected all the events with 
reconstructed direction within $10^{\circ}$ from the Earth limb, thus reducing contamination from 
Earth's $\gamma$-ray albedo.The total effective observation time, after the selection is 268 hours.
PG\,1553+113, observed at about 50 degrees off-axis with respect to the boresight, was not detected 
by the GRID at a significance level $>$ 3\, $\sigma$ and therefore an upper limit at 95$\%$ 
confidence level was calculated. Considering that AGILE has a higher particle background at very 
high off-axis angles, we calculated also the upper limit selecting only photons with E $\>$200\,MeV 
in order to minimize the possible contamination at low energies. The summary of the AGILE observations 
and the results of the analysis are reported in Table 1. During March-April 2008, PG\,1553+113 was 
outside the field of view of SuperAGILE, the hard X-ray (20-60 keV) imager on board AGILE 
\cite{Feroci2007}, thus no SuperAGILE data on this source are available from this period.

\section{X-rays: RXTE/ASM Observations}

The All Sky Monitor (ASM) on board the {\it Rossi} X-ray Timing Explorer ($R$XTE) satellite consists 
of three wide angle scanning shadow cameras. The cameras, mounted on a rotating drive assemble 
can cover almost $70 \%$ of the sky in every 1.5 hours \cite{Levine1996}. For this campaign, we 
used ASM data taken from  1 March through 31 May, 2008. The mean measured flux of PG\,1553+113 
is showed in Table 2.

\section{Optical data}

\subsection{KVA observations}

The KVA (Kungliga Vetenskaps Akademien) telescope is located at the Roque de los Muchachos 
Observatory on La Palma island, Spain and is operated by Tuorla Observatory.  
The telescope is composed of a 0.6\,m f/15 Cassegrain devoted to polarimetry, and a 0.35\,m 
f/11 SCT auxiliary telescope for multicolour photometry. This telescope has been successfully 
operated remotely since autumn 2003. The KVA is used for optical support observations for 
MAGIC by making $R$-band photometric observations typically one measurement per night per source. 
Photometric measurements of PG\,1553+113 were made in differential mode, i.e. by obtaining CCD 
images of the target and calibrated comparison stars in the same field of view 
(\cite{FiorucciTosti1996}; \cite{Fiorucci1998}; \cite{Villata1998}).

\subsection{REM observations}

REM (Rapid Eye Mount: a fast slewing robotized infrared telescope, \cite{covino}) acquired 
photometry of PG\,1553+113 on April 18 and 25, and May 2, 2008 with all the available filters 
($VRIJHK$).  The data reduction followed standard  procedures \cite {Dolcini2005}. The 
resulting flux of the three observations is reported in Table 2.  The optical magnitudes 
have been calibrated against the standard Landolt field \cite{landolt} of PG\,1323, while the 
near infrared magnitudes were calibrated against the 2MASS catalog. The optical data taken 
during the night of 18 April are affected by moonlight and have not been used. For the 
purpose of SED construction, all magnitudes have been dereddened using the dust IR maps 
\cite{Schlegel1998}.

\section{ Results}

Analysing the VHE data, an excess of 415 gamma-like events, over 1835 normalized background 
events was found, corresponds to a significance of 8\,$\sigma$. The differential VHE photon 
spectrum of PG\,1553+113, resulting from MAGIC data averaged over all observation nights, is 
plotted in Fig.1 (filled circles). It can be well described by a power law:
\begin{eqnarray}
{dN \over dE} = (2.0\pm0.3) \cdot \Big( {E \over 200 \mathrm{GeV}} \Big) ^{(-3.4\pm0.1)} 
\end{eqnarray}
in units of $10^{-10}$cm$^{-2}$s$^{-1}$TeV$^{-1}$ with $\chi^2$/DOF$=$1.36/3.  

During the nights of MAGIC observations no significant flux variability was found. The 
differential energy spectrum for the March data alone can be well described by:
\begin{eqnarray}
{dN \over dE} = (1.9\pm0.4) \cdot \Big( {E \over 200 \mathrm{GeV}} \Big)^{(-3.5\pm0.2)} \,,
\end{eqnarray}
whereas the energy spectrum for the April data can be well described by:
\begin{eqnarray}
{dN \over dE} = (2.1\pm0.4) \cdot \Big( {E \over 200 \mathrm{GeV}} \Big)^{(-3.3\pm0.2)} 
\end{eqnarray} 
(same units as above). Because of the interaction of VHE $\gamma$-rays with the extragalactic 
background light (EBL: the integrated light from all stars that ever formed; it spans the IR-UV 
frequency range), which leads to $e^+e^-$ pair production and ensuing attenuation of the VHE 
$\gamma$-ray flux, we computed the deabsorbed (i.e., intrinsic) fluxes using a specific 'low 
star formation model' of the EBL \cite{Kneiske2004}, assuming a source redshift of $z=0.3$. 
The resulting deabsorbed points are represented as empty squares in Fig.1.


 \begin{figure}[!t]
  \centering
  \includegraphics[width=2.5in]{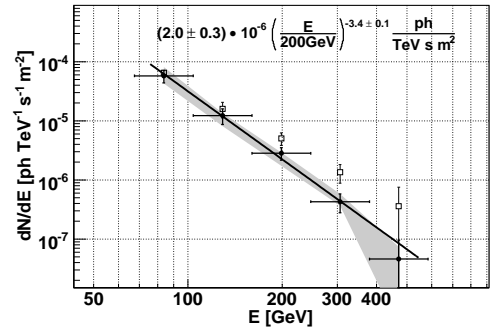}
\vspace{-0cm}  
\caption{The MAGIC measured spectrum of PG\,1553+113 (filled circles). 
The EBL-corrected points are shown as empty squares.)}
  \label{simp_fig}
 \end{figure}

The HE data reduction results from AGILE are summarized in Table 1. The upper limits obtained 
by AGILE are consistent with the average flux observed by \textit{Fermi}-LAT for this source 
in its first three months of operation \cite{Abdo2009}. The upper limit obtained in the second 
period (25-30 March) is of the same order of that obtained in the first period.

\begin{table}[!h]
\caption{The upper limit calculated by AGILE data in two different periods.}
\label{table:spec}
\centering
\begin{tabular}{|c|c|c|}
\hline
Time period &  Energy &  U.L. Flux [photons cm$^{-2}$s$^{-1}$] \\ \hline

 {16-21 March} &  $>$ 100MeV & $56 \times 10^{-8}$ \\
& $>$ 200 MeV &  $36 \times 10^{-8}$  \\ \hline
 {10-30 April } &   $>$ 100MeV & $34 \times 10^{-8}$ \\
& $>$ 200 MeV &  $21 \times 10^{-8}$  \\ \hline
\end{tabular}

\end{table}

The fluxes, and their corresponding effective photon frequencies, from 
the other telescopes used in this paper are reported in Table 2.

\begin{table}[!h]
\caption{Effective frequencies of operation, and corresponding fluxes 
from PG\,1553+113, from the various telescopes used in this paper.}
\label{table:spec}
\centering
\begin{tabular}{|c|c|c|}
\hline
Instrument & log($\nu$ [Hz]) &  log($\nu$F($\nu$) [erg cm$^{-2}$ s$^{-1}$]) \\ \hline
KVA     &    14.63   &  -10.2  \\ \hline
 {REM} &   14.38 & -10.33 \\
&14.27 & -10.34 \\
&14.13 & -10.38 \\ \hline
RXTE/ASM & 18.03 & -10.3 \\ \hline
\end{tabular}

\end{table}

\section{ Discussion}

The SED of PG\,1553+113 is shown in Fig.2. The VHE and HE $\gamma$-ray points are from MAGIC 
and AGILE, repectively. The X-ray point, provided by RXTE/ASM, represents the average flux 
between March 1 and May 31, 2008. The optical $R$-band point, provided by the KVA telescope, 
is from March 18,19. 

In addition to these  data, we also used an optical flux on April 18  from REM. To assess the 
soundness of the addition of this quasi-simultaneous data with other simultaneous data, we 
checked the optical variability of the source from April 13 to April 18 using KVA data, and 
found that the source was essentially stable, with a variability of less than 2$\%$ in flux.
In order to see the variability in HE range from the upper limits of AGILE, we used flux points 
from the \textit{Fermi} Gamma-ray Space Telescope (formerly GLAST; Abdo et al. 2009).


\begin{figure}[!t]
\centering
\vspace{-1cm} 
\includegraphics[width=2.5in]{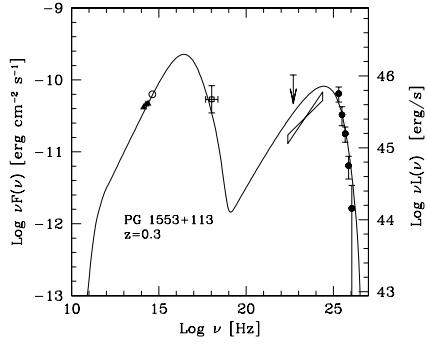}
\vspace{-2cm} 
\caption{The average SED of PG\,1553+113. The filled triangles denote the REM data. The open 
circle represents the KVA data point. The open square is from $R$XTE/ASM. At HE $\gamma$-rays, the 
upper limit is from AGILE. The filled circles show the deabsorbed MAGIC data points. We also 
show the non-simultaneous flux from {\it Fermi} (butterfly symbol).}
\label{simp_fig}
\end{figure}

We fit the resulting SED with a homogeneous one-zone SSC model \cite{Tavecchio2001}. In the 
frame of this model, the source is a spheric blob of plasma of radius $R$, moving with a 
Doppler factor $\delta$ \cite{Moriconi2006} towards the observer at an angle $\theta$ with 
respect to the line of sight and threaded with a tangled uniform magnetic field, $B$. The 
injected relativistic particle population is described as a broken power-law spectrum with 
normalization $K$, extending from $\gamma_{\rm min}$ to $\gamma_{\rm max}$ and with indices 
$n_{\rm 1}$ and $n_{\rm 2}$ below and above the break Lorentz factor $\gamma_{\rm br}$. In 
the framework of this model, the SED can be described with the following set of parameters: 
$\gamma_{\rm min}$ = 1, 
$\gamma_{\rm break} = 3 \times 10^4$, 
$\gamma_{\rm max} = 2 \times 10^5$, 
$K=0.5 \times 10^4$ cm$^3$, 
$n_{\rm 1}=2$, 
$n_{\rm 2}=4$, 
$B=0.7\,$G, 
$R=1.3 \times 10^{16}$ cm, and 
$\delta=23$. 

The {\it Fermi} and lowest-energy MAGIC data points do suggest some variability at HE $\gamma$-rays. 
Comparing the current SED with that based on earlier MAGIC observations \cite{Albert2007}, some 
variability is clearly visible. This is due to a small flux variation in the X-ray range and a large 
flux variation in the VHE range. The optical flux does not show any significant variability. 

Our results suggest that the variablility of PG\,1553+113 at different frequencies is highly time 
depended: hence, only a simultaneous multi-$\lambda$ monitoring over a large time span will give 
more information on the source. Relative to this fact, it is worth mentioning that the AGILE and 
MAGIC data presented here constitute the first simultaneous broad-band $\gamma$-ray observation 
(and ensuing SED) of any blazar.

\smallskip

{\it Acknowledgements.} The MAGIC collaboration would like to thank the Instituto de Astrof{\'\i}sica de Canarias for 
the excellent working condition at the Observetorio del Roque de los Muchachos at La Palma. 
Major support from Germany's Bundesministerium f\"ur Bildung, Wissenschaft, Forschung und 
Technologie and Max-Planck-Gesellschaft, Italy's Istituto Nazionale di Fisica Nucleare (INFN) 
and Istituto Nazionale di Astrofisica (INAF), and Spain's Ministerio de Ciencia e Innovaci{'o}n 
is gratefully acknowledged. The work was also supported by Switzerland's ETH Research grant 
TH34/043, Poland's Ministertwo Nauki i Szkolnictwa Wy$\dot{\rm z}$szego grant N N203 390834, and 
Germany's Young Investigator Program of the Helmholtz Gemeinschaft.

\end{document}